\documentclass[aps,prb,amsmath,amssymb,footinbib,showpacs,twocolumn,superscriptaddress,10pt]{revtex4-2}
\usepackage{amsmath}
\usepackage{amssymb}
\usepackage{amsthm}
\usepackage{setspace}
\usepackage{graphicx}
\usepackage{braket}
\usepackage{mathrsfs}
\usepackage{xcolor}
\usepackage{float}
\definecolor{ao(english)}{rgb}{0.0, 0.5, 0.0}
\definecolor{mygreen}{rgb}{0.29, 0.5, 0.1}
\usepackage[colorlinks = true,linkcolor = mygreen,urlcolor  = blue,citecolor = blue,anchorcolor = blue]{hyperref}
\usepackage[utf8]{inputenc}
\usepackage[english]{babel}
\usepackage{bm}
\usepackage{csquotes}
\MakeOuterQuote{"}
\usepackage{placeins}
\usepackage{comment}
\usepackage{multirow}
\usepackage{booktabs}

\begin{document}

\title{Majorana polarization in disordered heterostructures}
\author{Shubhanshu Karoliya}
\affiliation{School of Physical Sciences, Indian Institute of Technology Mandi, Mandi 175005, India.}
\author{Sumanta Tewari}
\affiliation{Department of Physics and Astronomy, Clemson University, Clemson, South Carolina 29634, USA}
\author{Gargee Sharma}
\affiliation{Department of Physics, Indian Institute of Technology Delhi, Hauz Khas, New Delhi 110016, India}

\begin{abstract}
Recent studies have proposed Majorana polarization as a diagnostic tool for identifying topological Majorana bound states (MBS) in engineered $p$-wave heterostructures.
In this work, we analyze the behavior of Majorana polarization in two systems: a one-dimensional (1d) semiconducting nanowire and a quasi-one-dimensional semiconducting system, both subject to Rashba spin-orbit coupling, proximity-induced superconductivity, and disorder. Our study reveals that Majorana polarization, while indicative of topological features, does not always distinguish true topological Majorana bound states from the partially separated Andreev bound states (psABS) or the quasi-Majorana modes, both in a 1d semiconducting nanowire as well as the quasi-1d system.
While earlier literature proposes that true topological MBS must satisfy the condition $\mathcal{P}^*_\mathrm{left} \cdot \mathcal{P}_\mathrm{right} \sim -1$, where $\mathcal{P}_\mathrm{left}$ and $\mathcal{P}_\mathrm{right}$ denote the Majorana polarizations in the left and right halves of the wire, it is well-established that two additional criteria must also be met for the identification of topological MBS and their applicability in topological quantum computation (TQC): the presence of a topological bandgap and the localization of wavefunctions at the edges.
We identify scenarios where the condition $\mathcal{P}^*_\mathrm{left} \cdot \mathcal{P}_\mathrm{right} \sim -1$ is satisfied without fulfilling the two additional conditions mentioned above. Our conclusions remain robust irrespective of the chosen definition of Majorana polarization, whether based on the chiral or the particle-hole framework.
We emphasize that robust identification of topological MBS useful for TQC requires combining diagnostics like polarization, energy spectra, and wavefunction localization, especially in disordered systems.
\end{abstract}

\maketitle

\section{Introduction}
Majorana fermions (MFs), originally proposed nearly a century ago as real solutions to the Dirac equation~\cite{Majorana1937}, hold a pivotal place in theoretical physics~\cite{das2023search}. While their existence as fundamental particle remains a profound question that is being investigated in several particle physics experiments, MFs have experienced a surprising and remarkable resurgence in condensed matter systems, where they emerge as excitations of complex many-body ground states and exhibit non-Abelian exchange statistics—a property with no counterpart in high-energy physics~\cite{moore1991nonabelions,read2000paired,nayak19962n}. Although fundamentally exotic, Majorana fermions are particularly compelling due to their potential applications in realizing fault-tolerant topological quantum computation (TQC)~\cite{Kitaev2003,Nayak2008}. A simple recipe to engineer MFs in solids is the following: a spin-orbit coupled semiconductor with induced superconductivity along with an externally applied Zeeman field should generate Majorana bound states (MBS) localized at the edges~\cite{sau2010generic, sau2010non,oreg2010helical,lutchyn2010majorana}. These states can then be braided to potentially realize topologically protected quantum gates and qubit operations.
Since the inception of this idea, intense theoretical and experimental efforts have sought to identify these quasiparticles in solids, though with only partial success~\cite{mourik2012signatures,deng2012anomalous,das2012zero,rokhinson2012fractional,churchill2013superconductor,finck2013anomalous,deng2016majorana,zhang2017ballistic,chen2017experimental,nichele2017scaling,albrecht2017transport,o2018hybridization,shen2018parity,sherman2017normal,vaitiekenas2018selective,albrecht2016exponential,Yu_2021,microsoft2025interferometric,mondal2025distinguishing,glodzik2020measure,rossi2020majorana,tanaka2024theory,tanaka2011symmetry,sharma2016tunneling,tanaka2009manipulation}.

One of the most extensively studied platforms for realizing MFs is a Rashba spin-orbit coupled one-dimensional (1d) semiconductor nanowire. Proximity-induced superconductivity and a Zeeman field applied parallel to the wire drive the system into a topologically nontrivial phase. Majorana bound states protected by a topological gap are localized at the ends of the nanowire and give rise to a quantized zero-bias conductance peak of height $2e^2/h$ in local differential conductance ($dI/dV$) measurements~\cite{sengupta2001midgap,law2009majorana, flensberg2010tunneling}. The quantization remains robust against variations in control parameters like magnetic field and gate voltage, leading to conductance plateaus. However, these features are also mimicked by topologically trivial states~\cite{Brouwer2012,Mi2014,Bagrets2012,pikulin2012zero,ramon2012transport,  pan2020physical,moore2018two,moore2018quantized,vuik2018reproducing,Stanescu_Robust,ramon_Jorge2106exceptional,ramon2019nonhermitian,Jorge2019supercurrent, sharma2020hybridization, zeng2020feasibility, zeng2022partially,
ramon2020from,  Jorge2021distinguishing, zhang2021, DasSarma2021,Frolov2021}. Specifically, in disordered systems, topologically unprotected Andreev bound states can emerge that appear very much like the MBSs. These are dubbed as partially separated Andreev bound states (psABS) or quasi-Majorana modes~\cite{moore2018two,moore2018quantized}. Here, due to the partial separation of the two Majorana components, local $dI/dV$ measurements, which are sensitive to only one of the modes, can mimic the signatures of topological MBS. It is thus crucial to employ additional diagnostics to confirm the topological origin of Majorana bound states.

Recently, Majorana polarization ($\mathcal{P}$) has been proposed as a reliable topological order parameter that can distinguish topological MBS from trivial zero-energy states~\cite{sticlet2012spin,sedlmayr2015visualizing,sedlmayr2016majorana,bena2017testing,kaladzhyan2017formation,awoga2024identifying}. A fermionic mode can be thought of as a superposition of two different Majorana modes (say $\chi_A$ and $\chi_B$), and Majorana polarization can be defined as just the difference between the probabilities of obtaining a $\chi_A$ mode and $\chi_B$ mode~\cite{sticlet2012spin}. While this definition is perfectly valid for systems only with chiral symmetry, in another work, Sedlmayr \textit{et al.}~\cite{sedlmayr2015visualizing} introduced a more general definition of Majorana polarization valid in systems with or without chiral symmetry. The Bogoliubov-de Gennes Hamiltonian ($H_\mathrm{BdG}$), which describes excitations of a superconductor, inherently has built-in particle-hole symmetry, which can be represented by a particle-hole operator $\mathcal{C}$ that anticommutes with the Hamiltonian. Since a true Majorana bound state is an equal superposition of both particle and hole components and has zero energy, it is an eigenstate of both the Hamiltonian $H_\mathrm{BdG}$ and the particle-hole operator $\mathcal{C}$. In this way,  the Majorana Polarization $\mathcal{P}$ can be defined as the expectation value of the particle-hole operator $\mathcal{C}$.
When $\mathcal{P}=0$, this implies the presence of both Majorana components, which couple to form a \textit{full} fermion. When $\mathcal{P}=\pm 1$, this indicates the presence of only one of the Majorana modes. One can typically evaluate $\mathcal{P}$ for both the left ($\mathcal{P}_\mathrm{left}$) and right ($\mathcal{P}_\mathrm{right}$) halves of a one-dimensional (1d) nanowire, and define the following topological condition based on non-local correlations between the left and right half of the system: $\mathcal{P}^*_\mathrm{left}\cdot\mathcal{P}_\mathrm{right}=-1$~\cite{awoga2024identifying}. 
Whether or not this condition is sufficient to distinguish topological MBS from the quasi-Majoranas remains unclear. 

In this work, we investigate Majorana polarization in two systems of interest: a semiconducting one-dimensional Rashba-coupled nanowire, which is chiral symmetric~\cite{tewari2012topological}, and a quasi-one-dimensional system that breaks chiral symmetry~\cite{tewari2012topological2}; both systems subject to induced superconductivity, an external Zeeman field, and disorder. We find that Majorana polarization, $\textit{alone}$, cannot always reliably distinguish between a true topological MBS and a psABS. 
While earlier studies on a semiconducting 1d wire propose that true topological MBS satisfy $\mathcal{P}^*_\mathrm{left} \cdot \mathcal{P}_\mathrm{right} \sim -1$, it is well-established that a topological bandgap and edge-localization of both Majorana components are also required for their identification and relevance to TQC.
We reveal cases in disordered systems where $\mathcal{P}^*_\mathrm{left} \cdot \mathcal{P}_\mathrm{right} \sim -1$ is satisfied but not the remaining two conditions, indicative of trivial Andreev bound states. Our conclusions hold consistently, regardless of whether Majorana polarization is defined within the chiral framework or the particle-hole framework. Robust identification of topological Majorana bound states necessitates a multifaceted approach combining polarization, energy spectra, and wavefunction localization, especially in disordered systems.

In the remainder of this paper, Section II discusses Majorana polarization in a 1D semiconducting nanowire, Section III explores Majorana polarization in a quasi-one-dimensional system, and conclusions are presented in Section IV.
\begin{figure*}
    \centering
    \includegraphics[width=1.9\columnwidth]{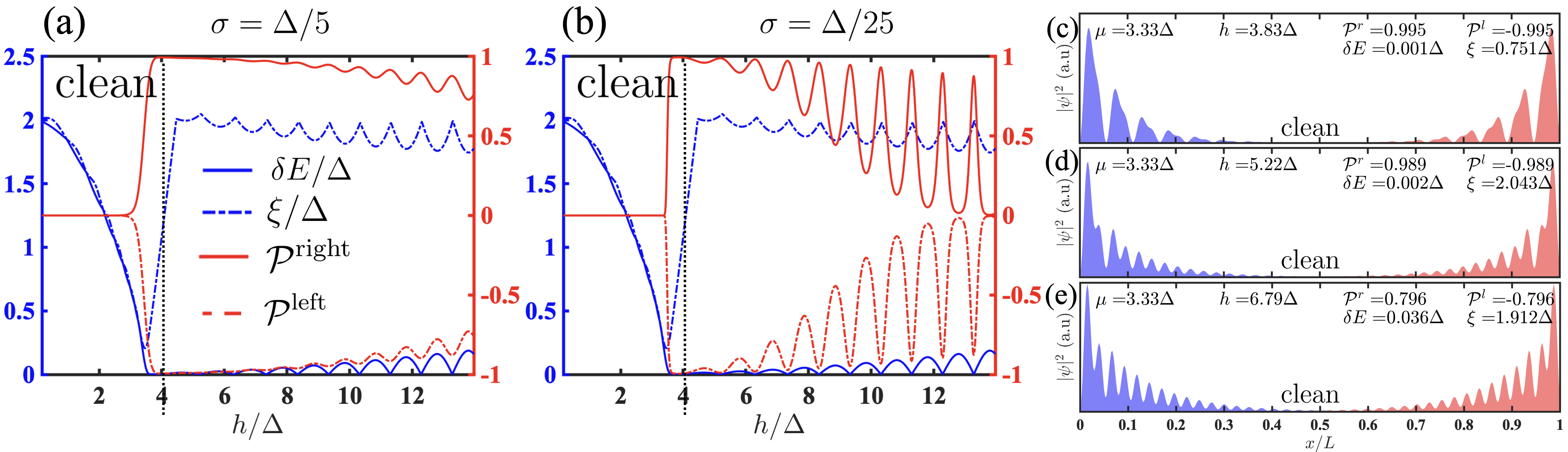}
    \caption{A clean one-dimensional nanowire. (a) Bandgap ($\xi$), the splitting between lowest energy modes ($\delta E$), the zero-energy Majorana polarization for both left ($\mathcal{P}^\mathrm{left}$) and right ($\mathcal{P}^\mathrm{right}$) half. The dashed black line separates the trivial and topological phases. We chose $\sigma=\Delta/5$. (b) Same as (a) but for a smaller Gaussian width $\sigma=\Delta/25$. A smaller Gaussian width makes the zero-energy polarization highly sensitive to the zero-energy splitting. (c)-(e) Majorana components $\chi_A$ and $\chi_B$ of the low-energy wavefunctions, showing localization of the MBS at the two ends of the wire--$\chi_A$ (light blue) is localized at one end of the wire and $\chi_B$ (light red). The relation between energy splitting and the Majorana polarization is also evident. }
    \label{fig:1d_nano_clean_majpol}
\end{figure*}

\section{Majorana polarization in a one-dimensional nanowire}
\subsection{Hamiltonian}
We begin with the following effective model that describes the low-energy physics of a one-dimensional semiconductor-superconductor hybrid structure:
\begin{align}
H &= -t\!\!\sum_{\langle i,j\rangle,\sigma}c_{i\sigma}^\dagger c_{j\sigma} + \sum_{i,\sigma}(V_\mathrm{dis}(x_i)-\mu) c^\dagger_{i\sigma}c_{i\sigma} \nonumber\\
&+h\!\sum_{i}c_i^\dagger \sigma_x c_i 
+\frac{\alpha_R}{2}\!\!\sum_{\langle i,j\rangle}\!\left(c_i^\dagger (i\sigma_y) c_j \!+\! \mathrm{h.c.}\right) \nonumber\\
&+\Delta\sum_i\left(c_{i\uparrow}^\dagger c_{i\downarrow} \!+\! \mathrm{h.c.}\right),  \label{Eq:Ham1D}
\end{align}
where $\langle i,j\rangle$ refer to the nearest-neighbor sites, the subscript $\sigma$ stands for spin, the column vector $c_i^\dagger = (c_{i\uparrow}^\dagger, c_{i\downarrow}^\dagger)$ is the electron creation operator on site $i$, and $\sigma_\nu$ (with $\nu=x, y, z$) are Pauli matrices. The parameters, $t$, $h$, $\mu$, $\alpha_R$, and $\Delta$ refer to the hopping parameters, external Zeeman field strength, chemical potential, Rashba spin-orbit coupling, and the induced superconducting pairing, respectively. $V_\mathrm{dis}(x_i)$ is the effective position-dependent disorder potential, which is modeled by considering $N_d$ randomly distributed short-range impurities. The potential profile of an isolated impurity located at position $x_i$ is 
\begin{equation}
V_\mathrm{imp}^{(i)}(x) = A_i \exp\left(-\frac{|x-x_i|}{\lambda}\right), \label{Eq:Vimp}
\end{equation}
where $\lambda$ is the characteristic decay length of the impurity potential and $A_i$ is a random amplitude characterized by a Gaussian probability distribution.  The net disorder potential is then given by 
\begin{equation}
V_\mathrm{dis}(x) = V_0 \sum_{i=1}^{N_d}A_i\exp\left(-\frac{|x-x_i|}{\lambda}\right),  \label{Eq:Vdis}
\end{equation}
where $V_0$ allows us to tune the overall magnitude of the disorder for a given disorder configuration. Unless otherwise specified, we use the following values for the parameters~\cite{zeng2022partially}: $a=5$nm is the lattice constant, $m=0.03 m_e$, where $m_e$ is the bare electronic mass, hopping parameter $t=50$meV ($t=\hbar^2/2ma^2$), $\alpha_R=250$meV$/a$, $\Delta = 0.3$meV, and $N=400$ is the number of lattice sites. The length of the wire is chosen to be 2$\mu$m.

\subsection{Majorana polarization: the chiral definition}
The Hamiltonian in Eq.~\ref{Eq:Ham1D} is chiral symmetric~\cite{tewari2012topological} and thus it suffices to use a chiral-symmetric definition of the Majorana polarization. We
recast the Hamiltonian in the Bogoliubov-de Gennes form in Nambu basis $(c^\dagger_{i\uparrow}, c^\dagger_{i\downarrow},c_{i\downarrow},-c_{i\uparrow})$ and then exactly diagonalize it numerically. 
Let us suppose a wavefunction solution of the Bogoluibov-de Gennes Hamiltonian is of the form $\psi_i(\epsilon) = (u_{i\uparrow}$, $u_{i\downarrow}$, $v_{i\downarrow}$, $v_{i\uparrow}$), where $\epsilon$ is the energy, and $u$ and $v$ correspond to the particle and hole components, respectively. Due to particle-hole symmetry, there also exists a negative energy solution $\psi_{i}(-\epsilon) = (v_{i\uparrow}^*,   v_{i\downarrow}^*, u_{i\downarrow}^*,   u_{i\uparrow}^*)$. We construct a new Majorana basis as follows:
\begin{align}
\gamma_{i\sigma}^A &= \frac{1}{\sqrt{2}}\left[c^\dagger_{i\sigma}+c_{i\sigma}\right], \nonumber\\
\gamma_{i\sigma}^B &= \frac{i}{\sqrt{2}}\left[c^\dagger_{i\sigma} -c_{i\sigma}\right].     \label{Eq_chi}
\end{align}
In the new basis ($\gamma^A_{i\uparrow}, \gamma^B_{i\downarrow}, \gamma^A_{i\downarrow}, \gamma^B_{i\uparrow}$), the wavefunction is expressed as 
\begin{align}
\psi_i = \frac{1}{\sqrt{2}}(u_{i\uparrow}-v_{i\uparrow}, -iu_{i\uparrow}-iv_{i\uparrow},u_{i\downarrow}-v_{i\downarrow}, -iu_{i\downarrow}-iv_{i\downarrow})
\end{align}
We define define the Majorana polarization $\mathcal{P}_i$ at site $i$ as the difference between the probability to have a $\gamma^A$
Majorana and the probability to have a $\gamma^B$ Majorana. Note that it is the difference of these two probabilities that gives that Majorana polarization, as their sum gives the probability of having both Majorana modes, which is equivalent to having a \textit{full} fermionic mode. 
The Majorana polarization $\mathcal{P}_i$ is thus evaluated to be~\cite{sticlet2012spin} 
\begin{align}
    \mathcal{P}_i &= -u_{i\uparrow}v^*_{i\uparrow} -v_{i\uparrow}u^*_{i\uparrow}+u^*_{i\downarrow}v_{i\downarrow}+ v^*_{i\downarrow}u_{i\downarrow}\nonumber\\
    &=2 \mathrm{Re}[u_{i\downarrow}v^*_{i\downarrow} - u_{i\uparrow}v^*_{i\uparrow}]
\end{align}
The local Majorana polarization at site $i$ and energy $\omega$ is thus defined to be: 
\begin{align}
    \mathcal{P}_i(\omega) = 2\sum_j \delta(\omega-\epsilon_j) \mathrm{Re}[u^j_{i\downarrow}v^{j*}_{i\downarrow} - u^j_{i\uparrow}v^{j*}_{i\uparrow}],
\end{align}
where the summation is over all the eigenvalues of the Hamiltonian. Numerically, we implement the Dirac-delta function by a narrow Gaussian function of width $\sigma$. One can then define the Majorana polarization in the left half of the wire as follows: 
\begin{align}
    \mathcal{P}_\mathrm{left}(\omega) = 2\sum_{i\leq N/2}\sum_j \delta(\omega-\epsilon_j) \mathrm{Re}[u^j_{i\downarrow}v^{j*}_{i\downarrow} - u^j_{i\uparrow}v^{j*}_{i\uparrow}].
\end{align}
Similarly, by summing from $N/2+1$ to $N$, one can define Majorana polarization in the right half of the wire-- $\mathcal{P}_\mathrm{right}(\omega)$. The product of $\mathcal{P}_\mathrm{right}$ and $\mathcal{P}_\mathrm{left}$ can measure correlations between the left and the right half of the wire. 
We conclude this subsection by emphasizing that the Majorana polarization is inherently a complex number ($\mathcal{P}=\mathcal{P}_x+i\mathcal{P}_y$) comprising both $x$ and $y$ components. However, for the 1d nanowire Hamiltonian discussed above, one component — specifically the $y$-component in the chosen coordinate system — vanishes. As such, $\mathcal{P}$ in this context refers explicitly to $\mathcal{P}_x$. We will revisit this discussion in the next section, where both components are nonzero in the case of quasi-1d system. In this case, the product of $\mathcal{P}_\mathrm{right}$ and $\mathcal{P}^*_\mathrm{left}$ turns out to be a useful quantity. 

\subsection{Majorana polarization in a clean 1d nanowire}
In Fig.~\ref{fig:1d_nano_clean_majpol}, we plot the Majorana polarization at $\omega=0$ averaged over one half of the system for a clean wire (no disorder) for both halves of the system. The polarizations $\mathcal{P}_\mathrm{left}$ and $\mathcal{P}_\mathrm{right}$ are antisymmetric to each other--they have equal magnitudes but are of opposite signs. As the magnetic field increases, the system undergoes a topological phase transition. The Majorana polarization rapidly increases in magnitude and acquires a maximum value in the topological region. For larger magnetic fields, the Majorana polarization shows \textit{oscillations of increasing magnitude}, similar to the oscillations of zero-energy splitting ($\delta E$)~\cite{sharma2020hybridization} and the bandgap ($\xi$), also shown in the figure. We find that the polarization is quite sensitive to the zero-energy splitting $\delta E$. When using a narrower Gaussian width ($\Delta/25$) to sample the polarization closer to $\omega=0$, the amplitude of the Majorana polarization oscillations increases significantly as seen in Fig.~\ref{fig:1d_nano_clean_majpol} (b). This is because the zero-energy splitting $\delta E$ is of the order of or greater than the Gaussian width $\sigma$, significantly reducing the magnitude of $\mathcal{P}$. 
\begin{figure}
    \centering
    \includegraphics[width=0.99\columnwidth]{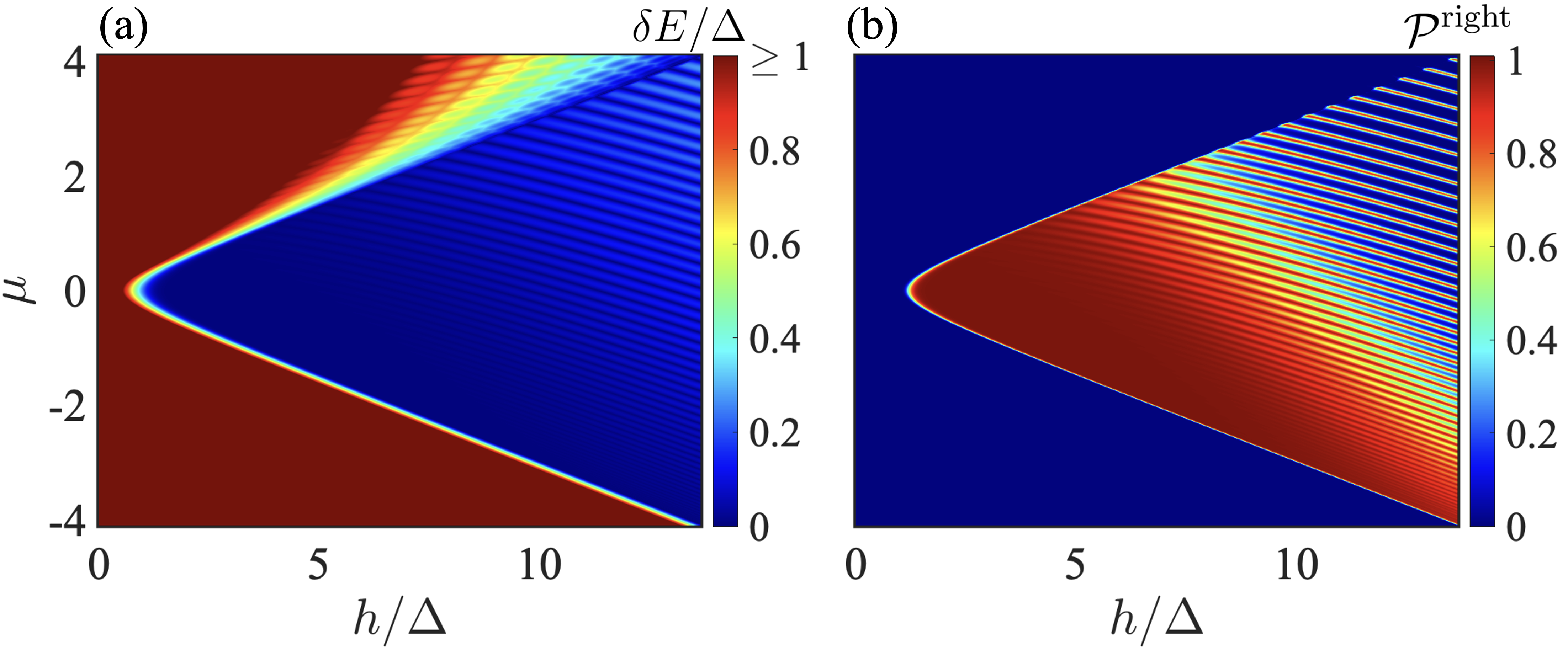}
    \caption{(a) The zero-energy splitting ($\delta E$) for a clean 1d nanowire plotted after exact diagonalization of Eq.~\ref{Eq:Ham1D} as a function of the Zeeman energy ($h$) and the chemical potential ($\mu$). (b) The corresponding Majorana polarization density $\mathcal{P}^\mathrm{right}$. We chose $\sigma=\Delta/25$.}
    \label{fig:1d_nano_clean_phase}
\end{figure}

For each pair of BdG eigenstates \( \psi_{\epsilon} \) and \( \psi_{-\epsilon} \), we can uniquely construct the modes \( \chi_A \) and \( \chi_B \), which correspond to the Majorana operators \( \gamma_A \) and \( \gamma_B \). This framework is particularly useful for studying low-energy excitations (\( \epsilon \ll \Delta \)) in Majorana nanowires. However, it is important to note that this construction is not limited to the low-energy regime but applies generally to all eigenstates of the BdG Hamiltonian.  
In Fig.~\ref{fig:1d_nano_clean_majpol}(c)-(e), we plot the probability distributions of the Majorana components \( \chi_A \) and \( \chi_B \) for the lowest-energy wavefunctions. In the topological phase, these Majorana bound states become localized at the ends of the nanowire, as illustrated in the figure. Importantly, $\chi_A$ (light blue) is localized at one end of the wire and $\chi_B$ (light red). In a scenario where both $\chi_A$ and $\chi_B$ are localized only near one end of the wire, they would not represent topological bound states. 
Furthermore, in the figure, the dependence of the polarization on the energy splitting \( \delta E \) is also visible, highlighting the interplay between the splitting and the spatial structure of the Majorana modes.
In Fig.~\ref{fig:1d_nano_clean_phase} (b) we plot the zero-energy $\mathcal{P}_\mathrm{right}$ as a function of the chemical potential and the Zeeman field strength. For comparison, we also plot the zero-energy splitting ($\delta E$) in Fig.~\ref{fig:1d_nano_clean_phase} (a). Both $\delta E$ and $\mathcal{P}_\mathrm{right}$ display oscillations in the topological region. The sensitivity of $\mathcal{P}_\mathrm{right}$ to oscillations in $\delta E$ is clearly evident at higher chemical potentials. 
\begin{figure}
    \centering
    \includegraphics[width=0.99\columnwidth]{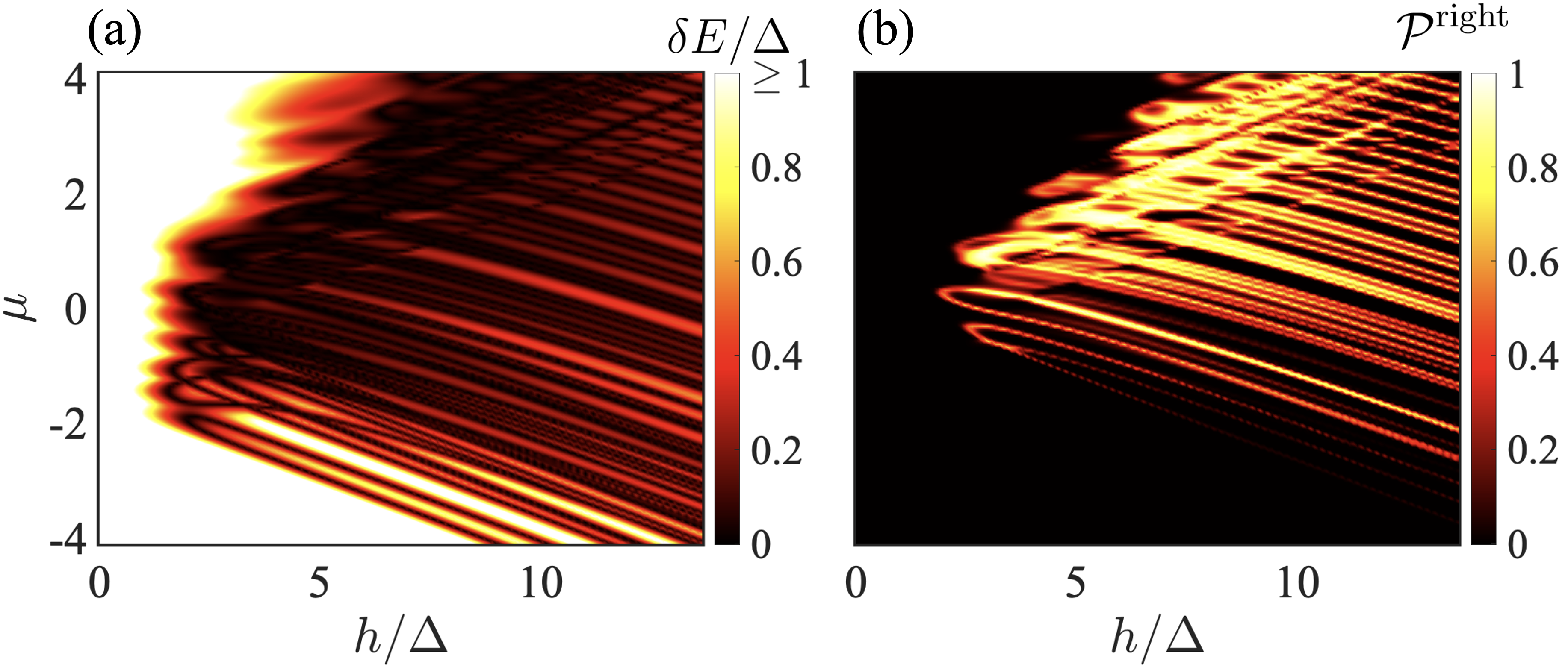}
    \caption{(a) The zero-energy splitting ($\delta E$) for a disordered nanowire plotted after exact diagonalization of Eq.~\ref{Eq:Ham1D} as a function of the Zeeman energy ($h$) and the chemical potential ($\mu$). (b) The corresponding Majorana polarization density $\mathcal{P}^\mathrm{right}$. We chose $\sigma=\Delta/25$. The disorder profile is shown in Fig.~\ref{fig:Vdis1:Pvsh} (b).}
    \label{fig:1d_dis_phase}
\end{figure}
\begin{figure}
    \centering
    \includegraphics[width=0.99\columnwidth]{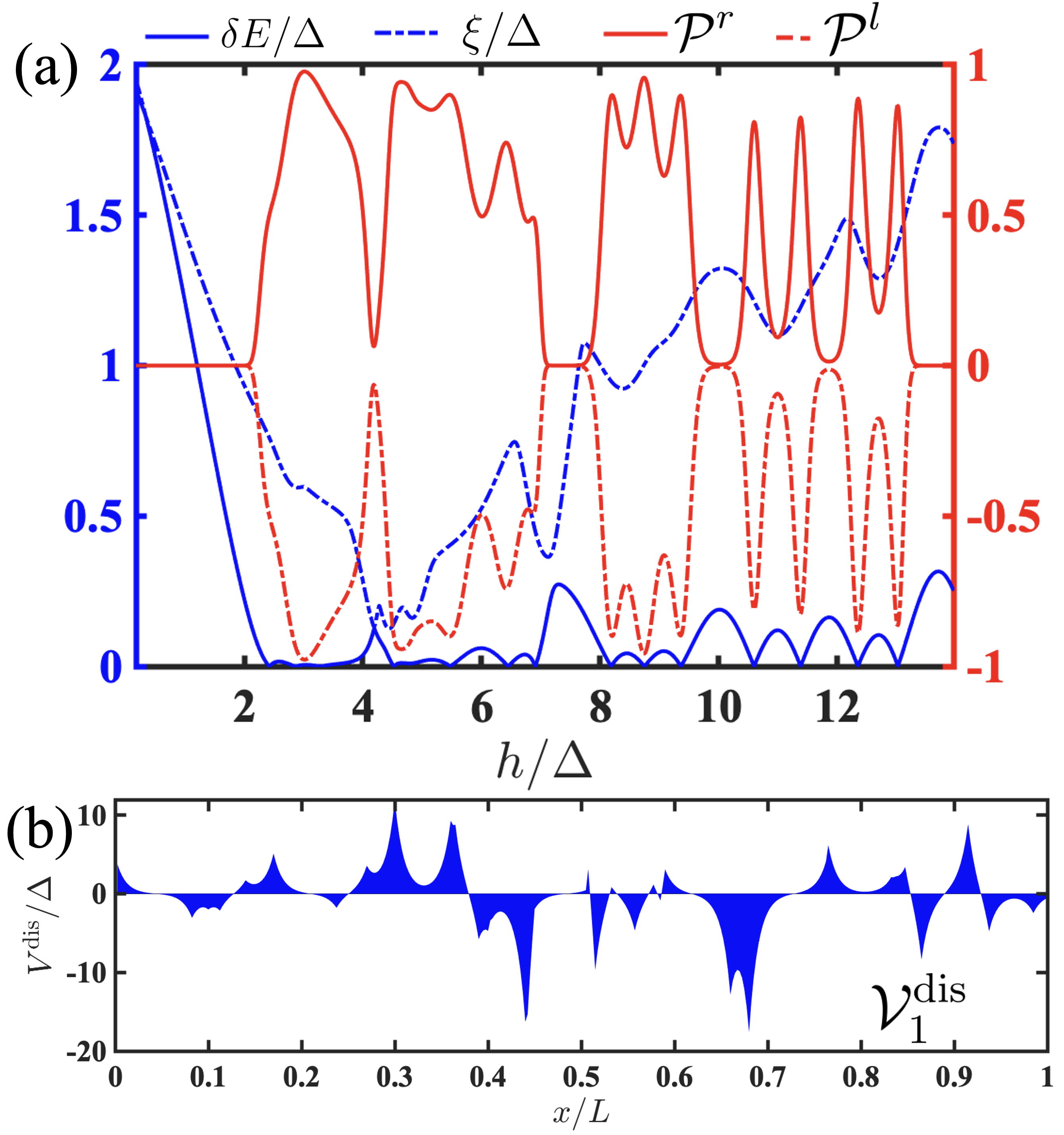}
    \caption{(a) Bandgap ($\xi$), splitting between lowest energy modes ($\delta E$), left ($\mathcal{P}^{l}$) and right ($\mathcal{P}^r$) Majorana polarization for a disordered one-dimensional nanowire as a function of applied Zeeman field $h$. We chose $\sigma=\Delta/25$. The product of the polarizations $\mathcal{P}^\mathrm{right}\cdot\mathcal{P}^\mathrm{left}\sim-1$ even in the trivial phase, i.e., when $h<h_c$, where the bandgap minimum occurs at $h_c\sim 4\Delta$.
    (b) The corresponding disorder profile denoted by $\mathcal{V}^\mathrm{dis}_1$ for the parameters $V_0=2$meV, $n_d=20/\mu$m, $\lambda=20$nm.}
    \label{fig:Vdis1:Pvsh}
\end{figure}

\begin{figure*}
    \includegraphics[width=2.05\columnwidth]{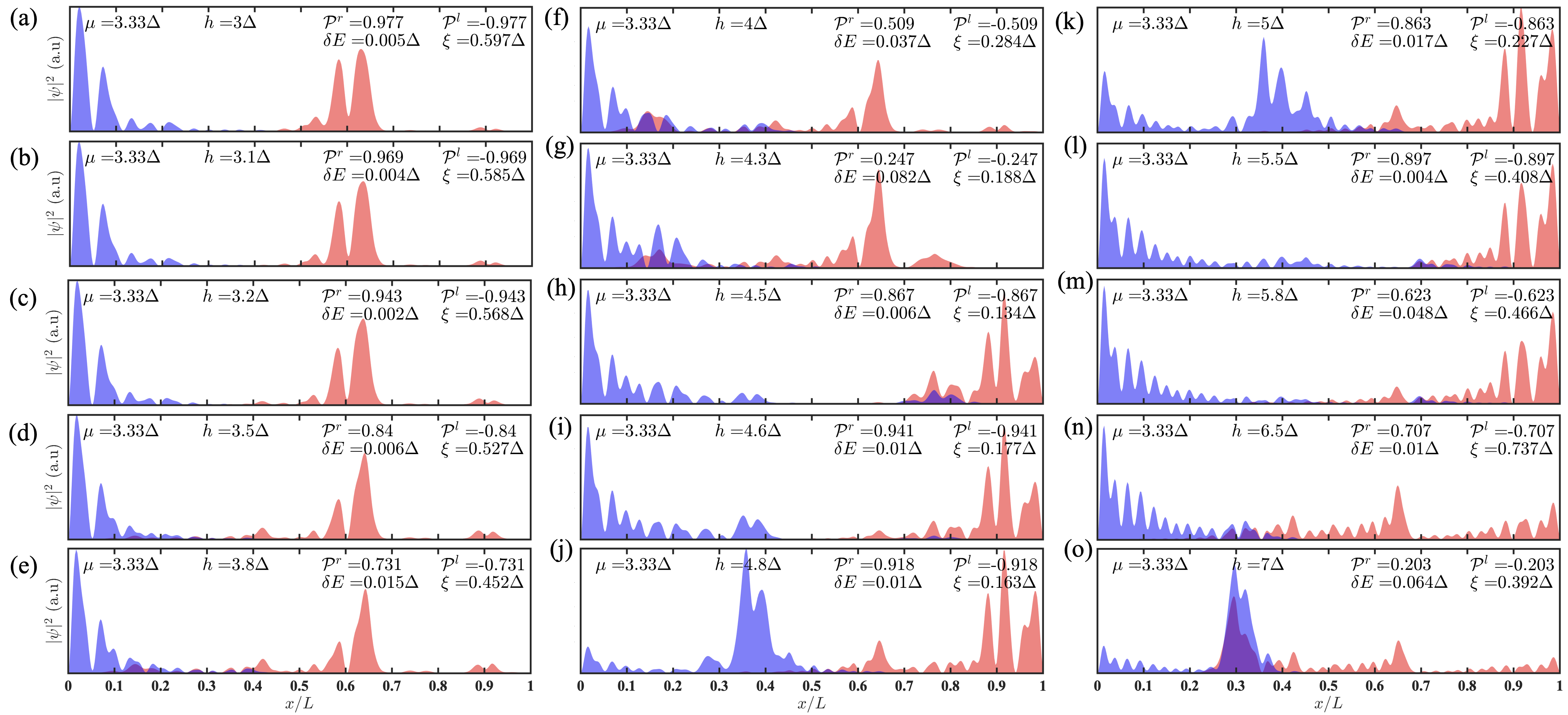}
    \caption{Majorana components of the lowest energy wavefunctions for a disordered one-dimensional nanowire. The chosen chemical potential $\mu$, applied Zeeman field $h$, and the obtained left ($\mathcal{P}^l$) and right ($\mathcal{P}^r$) Majorana polarization, along with the evaluated low-energy splitting $\delta E$ and bandgap $\xi$ is also shown in the inset. The emergence of partially separated Andreev bound states is clearly seen in (a)-(g) and (j). The chosen disorder configuration is $\mathcal{V}^{\mathrm{dis}}_1$ plotted in Fig.~\ref{fig:Vdis1:Pvsh} (b).}
    \label{Fig_1d_wavefunc_vdis1}
\end{figure*}
\subsection{Majorana polarization in disordered nanowires}
Having discussed Majorana polarization in a clean system, we now discuss the case of a disordered Majorana nanowire. In Fig.~\ref{fig:1d_dis_phase} (a) we plot the low-energy splitting ($\delta E$) for a disordered nanowire and in Fig.~\ref{fig:1d_dis_phase} (b) we plot the corresponding Majorana polarization $\mathcal{P}_\mathrm{right}$. While the clean system shows perfect correlations between $\delta E$ and $\mathcal{P}_\mathrm{right}$ (Fig.~\ref{fig:1d_nano_clean_phase}), this is not always the case with the disordered nanowire. For example, the oscillations in $\delta E $ around $\mu\sim -2$ and $h\sim 5\Delta$ are missing in the plot of $\mathcal{P}^\mathrm{right}$.
In Fig.~\ref{fig:Vdis1:Pvsh} (a), we plot the (i) bandgap $\xi$, (ii) the low-energy splitting ($\delta E$), (iii) left Majorana polarization ($\mathcal{P}^l$), and (iv) right Majorana polarization ($\mathcal{P}^r$) for a constant value of the chemical potential. The corresponding disorder profile, denoted by $\mathcal{V}^\mathrm{dis}_1$, is shown in Fig.~\ref{fig:Vdis1:Pvsh} (b), which was randomly generated using the parameters $V_0=2$meV, $n_d=20/\mu$m (impurity density), $\lambda=20$nm. 
From Fig.~\ref{fig:Vdis1:Pvsh} we note that the bandgap minimum occurs around $h(\equiv h_c)\sim 4\Delta$, which highlights the point of topological phase transition (note that the bandgap never fully closes due to finite size effects). Interestingly, the polarization product $\mathcal{P}^\mathrm{left}\cdot\mathcal{P}^\mathrm{right}\sim-1$ even in the trivial phase, i.e., when $h<h_c$.

In order to better understand the implications of Fig.~\ref{fig:Vdis1:Pvsh}, we plot the lowest energy Majorana wavefunctions in Fig.~\ref{Fig_1d_wavefunc_vdis1}. A generic feature that emerges is that the right Majorana polarization $\mathcal{P}^r$ is exactly negative of the left Majorana polarization $\mathcal{P}^l$ even in the presence of disorder.
Next, plots ~\ref{Fig_1d_wavefunc_vdis1} (a) to ~\ref{Fig_1d_wavefunc_vdis1} (g) clearly demonstrate the emergence of partially separated Andreev bound states (quasi-Majorana modes) in the trivial phase ($h<h_c$). Here, a single Majorana mode is localized at one end of the nanowire. While well separated with negligible overlap, the other mode is not localized at the opposite end of the wire and remains only partially separated from the first mode; in our case, a bit to the right of the middle of the wire. Interestingly, the magnitude of the Majorana polarization can approach unity, akin to the clean system. Moreover, as shown in Fig.~\ref{Fig_1d_wavefunc_vdis1} (d), (e), the Majorana polarization is highly sensitive to the overlap between the two modes. Thus, the partially separated Andreev bound states can exhibit a Majorana polarization close to one or even less than one (and subsequently the product of left and right polarization to be -1 or greater than -1), depending on the degree of overlap between the states. Fig.~\ref{Fig_1d_wavefunc_vdis1} (j), (k) also exhibit signatures of quasi-Majorana modes, supporting similar conclusions as discussed above.
Fig.~\ref{Fig_1d_wavefunc_vdis1} (h), (i), (l), and (m) highlight true topological Majorana bound states, localized at the two ends of the wire. Once again, the Majorana polarization magnitude is seen to vary in a broad range--from 0.6 to nearly 1. A key distinction from the clean system is that while in the clean case the overlap is proportional to the energy splitting of the modes, in the disordered case, this dependence is not necessarily linear. The polarization is proportional to the overlap rather than the energy splitting of the low-energy modes because the relation between energy splitting of the modes and the overlap is not monotonic for disordered systems.
These findings strongly suggest that Majorana polarization may not always reliably distinguish a true topological Majorana mode from a quasi-Majorana mode.

\section{Majorana polarization in a quasi-one-dimensional system}
\subsection{Hamiltonian}
We consider the following effective model that describes the low-energy physics of a generic two-dimensional semiconductor-superconductor hybrid structure:
\begin{align}
H &= -t\!\!\sum_{\langle \mathbf{r}_i,\mathbf{r}_j\rangle,\sigma}c_{\mathbf{r}_i\sigma}^\dagger c_{\mathbf{r}_j\sigma} + \sum_{\mathbf{r}_i,\sigma}(V_\mathrm{dis}(\mathbf{r}_i)-\mu) c^\dagger_{\mathbf{r}_i\sigma}c_{\mathbf{r}_i\sigma} \nonumber\\
&+h\!\sum_{\mathbf{r}_i}c_{\mathbf{r}_i}^\dagger \sigma_x c_{\mathbf{r}_i} 
+\frac{\alpha_R}{2}\!\!\sum_{\langle x_i,x_j\rangle}\!\left(c_{x_i}^\dagger (i\sigma_y) c_{x_j} \!+\! \mathrm{h.c.}\right) \nonumber\\
&-\frac{\alpha_R}{2} \!\!\sum_{\langle y_i,y_j\rangle}\!\left(c_{y_i}^\dagger (i\sigma_x) c_{y_j} \!+\! \mathrm{h.c.}\right)
+\Delta\sum_{\mathbf{r}_i}\left(c_{\mathbf{r}_i\uparrow}^\dagger c_{\mathbf{r}_i\downarrow} \!+\! \mathrm{h.c.}\right). \label{Eq:Ham2D}
\end{align}
Here $\mathbf{r}=(x_i,y_i)$ refers to the position coordinate, and $\langle ...\rangle$ refers to the nearest-neighbor sites. The hopping parameter $t$ and the Rashba spin-orbit coupling $\alpha_R$ now couple sites along both the $x$ and $y$-directions. 
The disorder potential $V_\mathrm{dis}(\mathbf{x}_i)$ is modeled similar to the 1D potential in Section II but now extends along the $y$-direction as well, i.e., we choose $N_d$ randomly distributed short-range impurities, and the potential profile of an isolated impurity located at position $\mathbf{x}_i$ is 
\begin{equation}
V_\mathrm{imp}^{(i)}(\mathbf{x}) = A_i \exp\left(-\frac{|\mathbf{x}-\mathbf{x}_i|}{\lambda}\right), \label{Eq:Vimp2d}
\end{equation}
The net disorder potential becomes
\begin{equation}
V_\mathrm{dis}(\mathbf{x}) = V_0 \sum_{i=1}^{N_d}A_i\exp\left(-\frac{|\mathbf{x}-\mathbf{x}_i|}{\lambda}\right).  \label{Eq:Vdis2d}
\end{equation} 
\begin{figure}
    \centering
    \includegraphics[width=1\columnwidth]{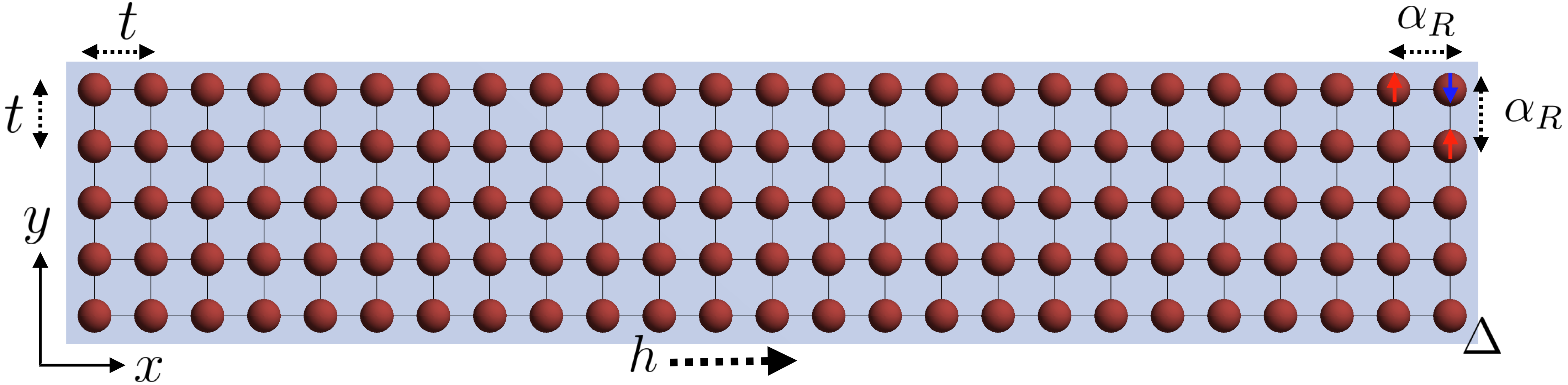}
    \caption{Schematic of a quasi-one-dimensional system. The labels $t$, $\alpha_R$, and $h$ refer to the spin-independent nearest neighbor hopping, the spin-dependent nearest neighbor hopping (due to Rashba spin-orbit coupling), and the applied external Zeeman field. The system is placed in the vicinity of a superconductor (highlighted blue) that induces a superconducting gap $\Delta$ through the proximity effect. }
    \label{fig:2dlattice}
\end{figure}
Fig.~\ref{fig:2dlattice} schematically shows the quasi-one-dimensional system. Unless otherwise specified, we use the following values for the parameters~\cite{zeng2022partially}: $a=5$nm is the lattice constant, $m=0.03 m_e$, where $m_e$ is the bare electronic mass, hopping parameter $t=50$meV ($t=\hbar^2/2ma^2$), $\alpha_R=250$meV$/a$, $\Delta = 0.3$meV, $l_x=2\mu$m is the length of the chain in the $x$-direction, and $l_y=.025\mu$m is the length of the chain along the $y$-direction. Thus, we have 2000 lattice sites, comprising five coupled chains with 400 lattice sites each, making it effectively a quasi-one-dimensional system. Note that we specifically need to choose an odd number of chains as the Majorana modes annihilate in pairs when the number of chains is even.

\subsection{Majorana Polarization: the particle-hole definition}
The Hamiltonian in Eq.~\ref{Eq:Ham2D} breaks chiral symmetry due to the transverse Rashba SOC term~\cite{tewari2012topological}. It is thus useful to employ a more general definition of Majorana polarization. 
In a recent work, Sedlmayr \textit{et al.}~\cite{sedlmayr2015visualizing} introduced a general definition of Majorana polarization valid in systems with or without chiral symmetry. The Bogoliubov-de Gennes Hamiltonian ($H_\mathrm{BdG}$), which describes excitations of a superconductor, inherently has built-in particle-hole symmetry. This symmetry operation is represented by a particle-hole operator $\mathcal{C}$ that anticommutes with the $H_\mathrm{BdG}$. Since a true Majorana bound state $|\psi\rangle$ is an equal superposition of both particle and hole components, it is an eigenstate of both the Hamiltonian $H_\mathrm{BdG}$ and the particle-hole operator $\mathcal{C}$, and therefore must occur at exactly zero energy ($[H_\mathrm{BdG},\mathcal{C}]_{\pm}|\psi\rangle=0$, where $\pm$ refer to commutator/anti-commutator).  The Hamiltonian is recast in the Nambu basis $(c^\dagger_{\mathbf{r}_i\uparrow}, c^\dagger_{\mathbf{r}_i\downarrow},c_{\mathbf{r}_i\downarrow},-c_{\mathbf{r}_i\uparrow})$ and then exactly diagonalized numerically. 
The wavefunction solution of the Bogoluibov de Gennes Hamiltonian is written the form $\psi(\mathbf{r}_i,\epsilon) = (u_{\mathbf{r}_i\uparrow}$, $u_{\mathbf{r}_i\downarrow}$, $v_{\mathbf{r}_i\downarrow}$, $v_{\mathbf{r}_i\uparrow}$), where $\epsilon$ is the energy, and $u$ and $v$ correspond to the particle and hole components, respectively. The particle-hole operator the takes the form $\mathcal{C} = \sigma_y\tau_y \mathcal{K}$, where $\sigma_y$ and $\tau_y$ are Pauli spin-$y$ matrices, and $\mathcal{K}$ represents complex conjugation. The local Majorana polarization $\mathcal{P}(\mathbf{r}_i,\epsilon)$ at a particular site $\mathbf{r}_i$ is defined to be the expectation value of the particle hole operator, i.e., $\mathcal{P}(\mathbf{r}_i,\epsilon_j)=\langle\psi(\mathbf{r}_i,\epsilon_j) | \mathcal{C}|\psi(\mathbf{r}_i,\epsilon_j)\rangle$. Furthermore, if an MBS is localized in a region $\mathcal{R}$ in space, the normalized polarization should be a complex number with unit magnitude, i.e., 
\begin{align}
\mathcal{P}_\mathcal{R}(\epsilon_j)=\frac{\sum_{i\in\mathcal{R}}\langle\psi(\mathbf{r}_i,\epsilon_j) | \mathcal{C}|\psi(\mathbf{r}_i,\epsilon_j)\rangle}{\sum_{i\in\mathcal{R}}\langle\psi(\mathbf{r}_i,\epsilon_j) | \psi(\mathbf{r}_i,\epsilon_j)\rangle} = e^{i\xi}.
\label{Eq:Pregion1}
\end{align}
To evaluate the polarization at a finite frequency $\omega$ we sum over the eigenstates weighted by a Dirac-delta function, i.e., 
\begin{align}
    \mathcal{P}_\mathcal{R}(\omega)=\frac{\sum_{i\in\mathcal{R}}\sum_j\langle\psi(\mathbf{r}_i,\epsilon_j) | \mathcal{C}|\psi(\mathbf{r}_i,\epsilon_j)\rangle \delta(\epsilon_j-\omega)}{\sum_{i\in\mathcal{R}}\sum_j\langle\psi(\mathbf{r}_i,\epsilon_j) | \psi(\mathbf{r}_i,\epsilon_j)\rangle \delta(\epsilon_j-\omega)}
    \label{Eq:P_region2}
\end{align}
The summation $j$ is over all the eigenstates, and the Dirac-Delta function is implemented by a narrow Gaussian factor of width $\sigma$. Unless otherwise specified, we choose $\sigma=\Delta/25$.
Note that $\mathcal{P}_\mathcal{R}$ is a complex number, with real and imaginary parts, $\mathcal{P}_\mathcal{R}=\mathcal{P}_\mathcal{R}^x+i\mathcal{P}_\mathcal{R}^y$. Unlike the strictly 1d nanowire in Sec. II, both the components here are found to be in general nonzero.
\begin{figure*}
    \includegraphics[width=2\columnwidth]{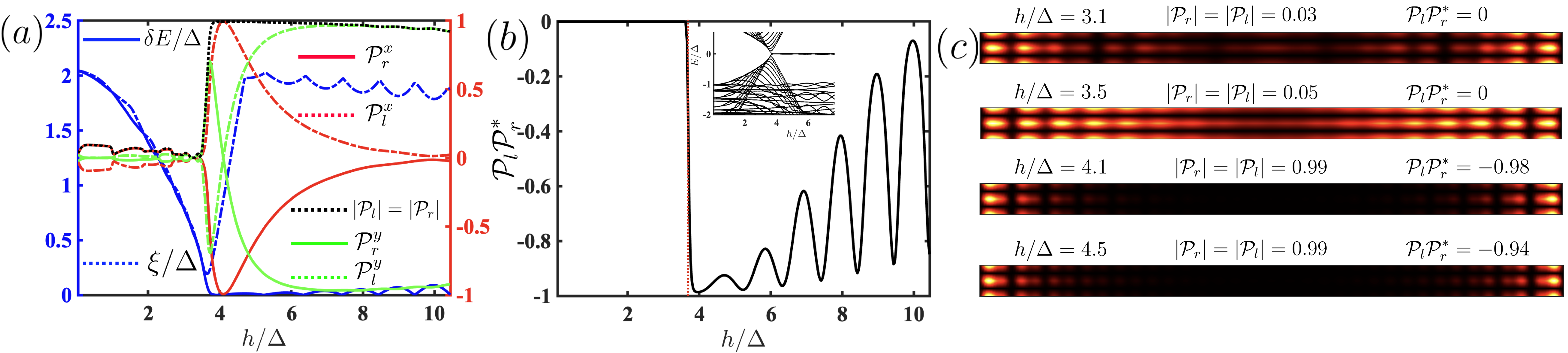}
    \caption{Quasi-1d system with five coupled chains described by the Hamiltonian in Eq.~\ref{Eq:Ham2D} with no disorder. (a) The bandgap ($\xi$), splitting between the lowest energy modes ($\delta E$), and the left ($l$) and right ($r$) polarization for the $x$ and $y$-components of the zero-energy polarization $\mathcal{P}$, which are shown by the red and green curves, respectively. The black dotted curve indicates the magnitude of $\mathcal{P}$, which is the same for both halves of the wire. 
    (b) The product $\mathcal{P}_l\mathcal{P}_r^*$ (which is equal to $\mathcal{P}_l^*\mathcal{P}_r$) as a function of the applied Zeeman field. The dashed red line indicates a topological phase transition identified by the bandgap minimum. The inset shows the low-energy quasiparticle spectrum.
    (c) Probability distribution of the low-energy Majorana components for different values of the Zeeman field.}
    \label{Fig_clean_quasi1d}
\end{figure*}
\begin{figure*}
    \includegraphics[width=2\columnwidth]{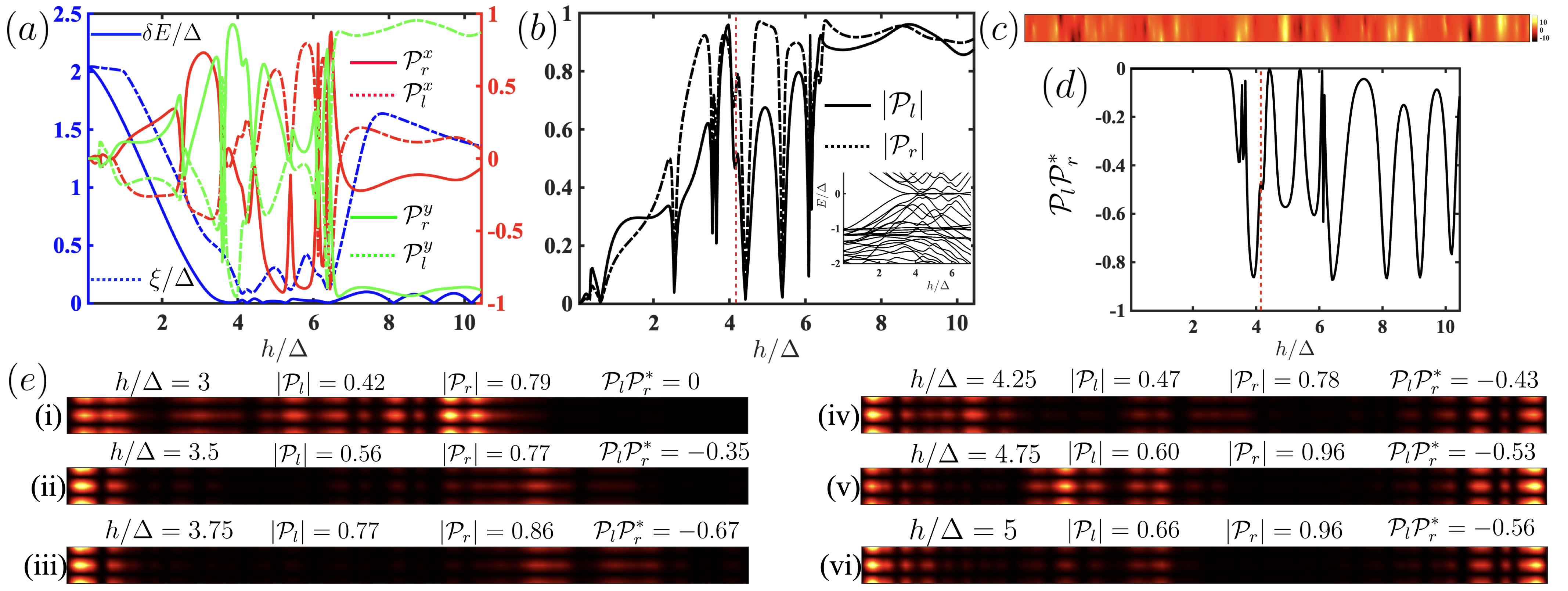}
    \caption{Quasi-1d system with five coupled chains described by the Hamiltonian in Eq.~\ref{Eq:Ham2D} in the presence of disorder. The bandgap ($\xi$), splitting between the lowest energy modes ($\delta E$), and the left ($l$) and right ($r$) polarization for the $x$ and $y$-components of the zero-energy polarization $\mathcal{P}$, which are shown by the red and green curves, respectively. (b) Magnitude of $\mathcal{P}$, which is now different for both halves of the wire. The dashed red line indicates a topological phase transition identified by the bandgap minimum. The inset shows the low-energy quasiparticle spectrum. (c) The disorder potential. 
    (d) The product $\mathcal{P}_l\mathcal{P}_r^*$ (which is equal to $\mathcal{P}_l^*\mathcal{P}_r$) as a function of the applied Zeeman field. The dashed red line indicates a topological phase transition identified by the bandgap minimum. (e) Probability distribution of the low-energy Majorana components for different values of the Zeeman field. As the Zeeman field is increased, one transitions from trivial to quasi-Majorana phase, and finally to the topological phase with true MBS at both ends of the system.}
    \label{Fig_disquasi1d_1}
\end{figure*}

The inset in Fig.~\ref{Fig_clean_quasi1d} (b) shows the quasiparticle spectrum of the Hamiltonian described in Eq.~\ref{Eq:Ham2D} in the absence of disorder. A neat topological phase transition is seen above the critical magnetic field (happens at $h_c \equiv h\sim 3.7\Delta$ in this case), and the low-energy Majorana modes appear. As usual, we divide our system into left and right halves and evaluate the zero-energy polarization $\mathcal{P}$ in each region using Eq.~\ref{Eq:P_region2}. We plot this in Fig.~\ref{Fig_clean_quasi1d} (a). All four quantities, $\mathcal{P}^x_l$, $\mathcal{P}^x_r$, $\mathcal{P}^y_l$, and $\mathcal{P}^y_r$ are close to zero in the trivial phase and acquire nonvanishing values after the topological phase transition. Close to the critical point (when $h$ is only slightly greater than $h_c$), the polarization has both nonvanishing $x$ and $y$-components, and at higher values of the magnetic field, only $\mathcal{P}^y$'s remain significant. Furthermore, the left and the right halves have equal and opposite values ($\mathcal{P}^x_l=-\mathcal{P}^x_r$ and $\mathcal{P}^y_l=-\mathcal{P}^y_r$). The magnitude $\mathcal{P}_l$ (which is the same as $\mathcal{P}_r$) is close to zero in the trivial phase and close to unity in the topological phase. 

Ref.~\cite{awoga2024identifying} recently introduced the quantity $\mathcal{P}_l^*\mathcal{P}_r$ as a measure of identifying non-local correlations between the two halves of the system.
In Fig.~\ref{Fig_clean_quasi1d} (b), we plot the product $\mathcal{P}_l^*\mathcal{P}_r$ (which is equal to $\mathcal{P}_r^*\mathcal{P}_l$) as a function of the applied Zeeman field. It remains exactly zero in the trivial phase and becomes nonzero in the topological phase.  Furthermore, as the Zeeman field varies in the topological phase, it oscillates between 0 and -1, and the amplitude of oscillations is seen to increase with the increase in the Zeeman field. Notably, increasing the Gaussian width $\sigma$ reduces the amplitude of these oscillations, consistent with the behavior observed earlier for the 1D nanowire. Suppose we instead plot $\mathcal{P}_l^* \mathcal{P}_r$ for only the lowest-energy mode, without the Dirac-delta weighting introduced in Eq.~\ref{Eq:P_region2}, the oscillations disappear, as shown in Fig.~\ref{fig:PlPr_compare_nogauss} (a). This occurs because, unlike in Fig.~\ref{Fig_clean_quasi1d} (b), we are no longer sampling the polarization specifically at zero energy.
As the Zeeman field is increased in Fig.~\ref{fig:PlPr_compare_nogauss} (a), the product $\mathcal{P}_l^* \mathcal{P}_r$ approaches a value close to -1 and then gradually decreases, exhibiting only small residual oscillations. The small residual oscillations arise due to the overlap between the two Majorana modes.
Fig.~\ref{Fig_clean_quasi1d} (c) shows the probability distribution of the low-energy Majorana components. The formation of localized modes at the two ends of the quasi-1d system is clearly seen in the topological phase. 

\subsection{Majorana polarization in disordered quasi-one-dimensional systems}
We next discuss the formation of partially separated Andreev bound states (quasi-Majoranas), topological Majorana bound states, and the behavior of Majorana polarization in the disordered quasi-1d system.  
The inset in Fig.~\ref{Fig_disquasi1d_1} (b) shows the low-energy quasiparticle spectrum a disordered configuration specified by the following parameters: $V_0=2$meV, $\lambda=10$nm, $n_d=0.05/$site. The zero energy modes are seen to appear before the closing of the topological gap (that happens at $h_c \equiv h\sim 4.1\Delta$ in this case), which is indicative of low-energy Andreev bound states that are topologically trivial. This is more clearly demonstrated in Fig.~\ref{Fig_disquasi1d_1} (a), where the dashed blue curve shows the bandgap, while the solid blue curve shows the lowest energy mode splitting. The quantities $\mathcal{P}^x_l$, $\mathcal{P}^x_r$, $\mathcal{P}^y_l$, and $\mathcal{P}^y_r$, plotted in Fig.~\ref{Fig_disquasi1d_1} (a), no longer vanish in the trivial phase (they were close to zero in the clean system). Furthermore, the magnitude of the polarizations $|\mathcal{P}_l|$ and $|\mathcal{P}_r|$ can be as big as unity in the trivial phase, and oscillates between 0 and 1 in the topological phase. This is also in sharp contrast to the clean system, where $|\mathcal{P}_l|$ and $|\mathcal{P}_r|$ were zero in the trivial phase and close to 1 in the topological phase. Also, $|\mathcal{P}_l|$ and $|\mathcal{P}_r|$ are no longer equal in magnitude, unlike in the clean system. Note that $|\mathcal{P}_l|$ and $|\mathcal{P}_r|$ were equal even in the disordered case for a 1d nanowire, and this distinction is because of the different definitions of Majorana polarizations--the chiral definition followed in Sec. II and the more general particle-hole definition (with normalization) followed in this section. 
\begin{figure}
    \centering
    \includegraphics[width=\columnwidth]{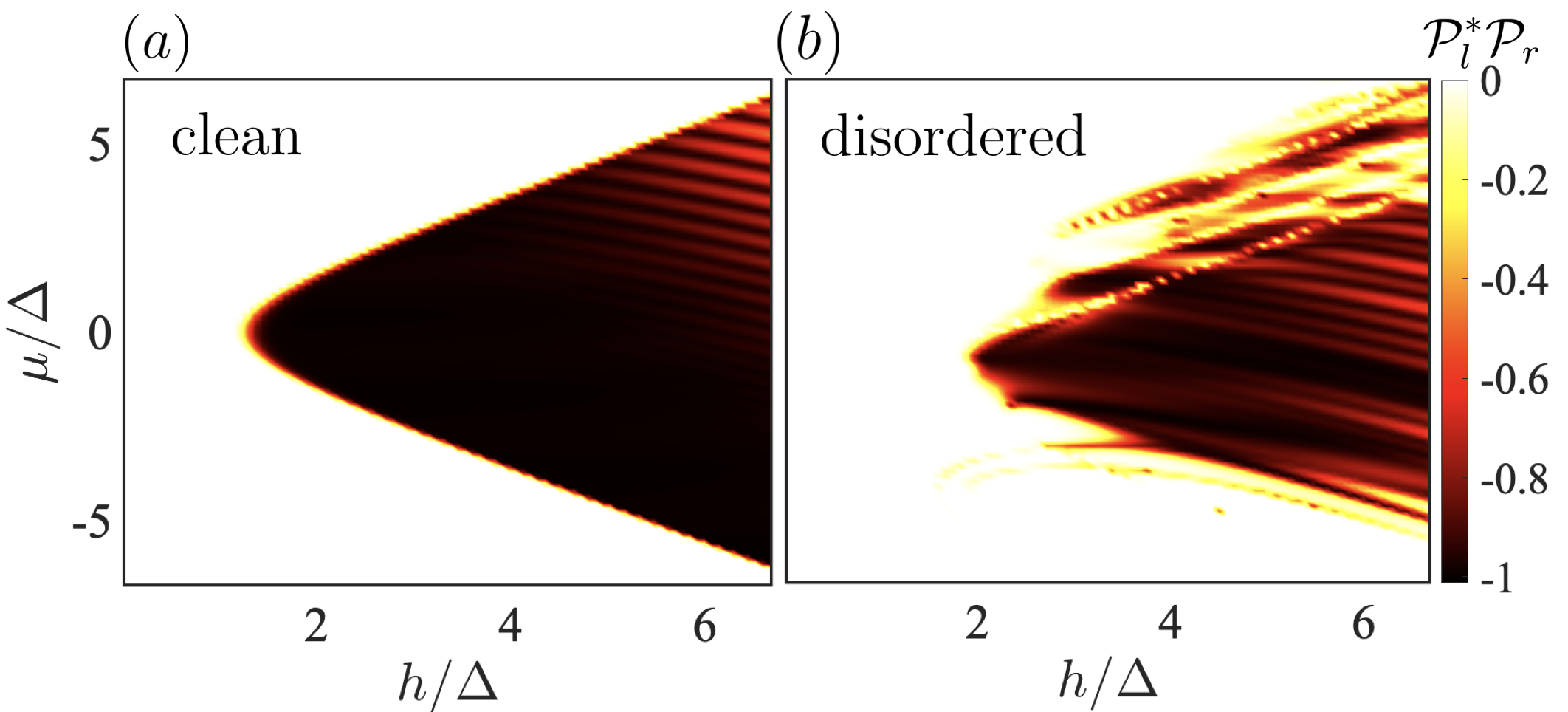}
    \caption{The product $\mathcal{P}_l^*\mathcal{P}_r$ for (a) clean and (b) disordered quasi-1d-system as a function of the Zeeman field and chemical potential.}
    \label{fig:plstarprphase}
\end{figure}
\begin{figure}
    \centering
    \includegraphics[width=\columnwidth]{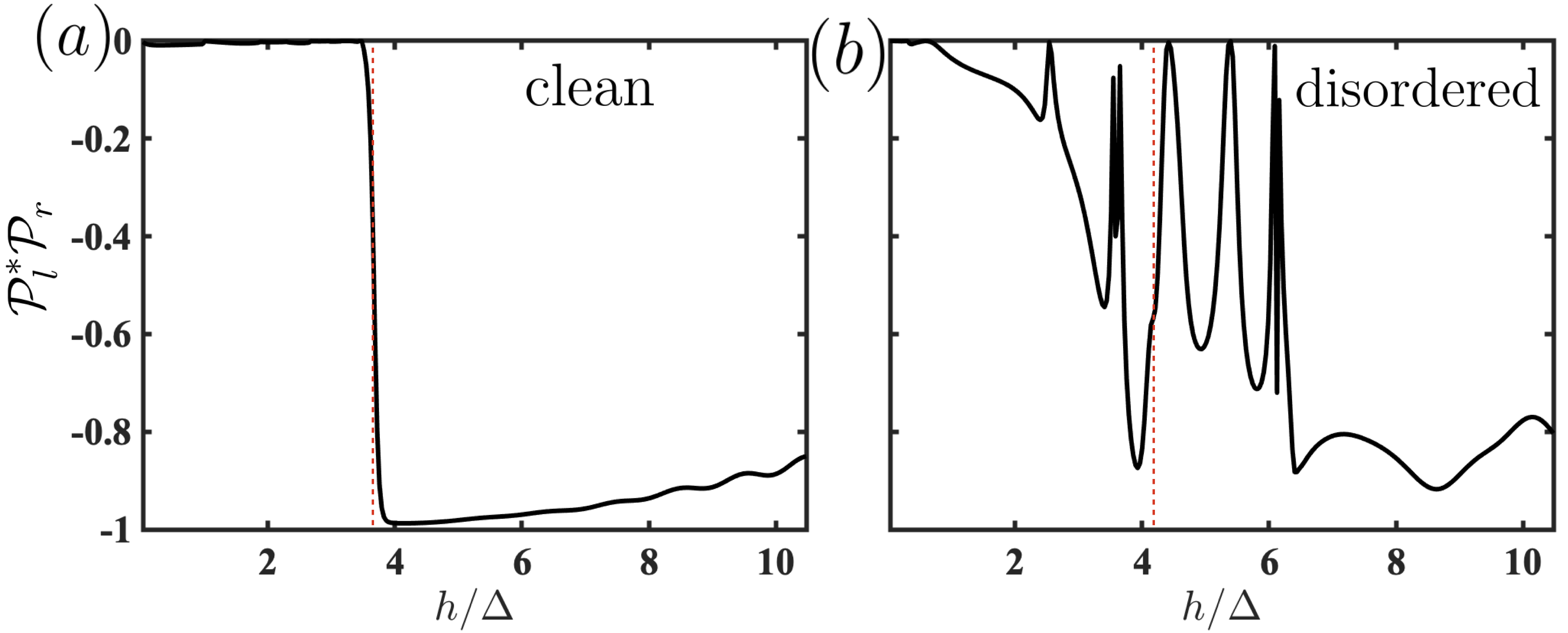}
    \caption{The product $\mathcal{P}_l^*\mathcal{P}_r$ for the lowest energy mode for the quasi-1d-system plotted without the Dirac-Delta function weighting introduced in Eq.~\ref{Eq:P_region2}. Dashed red lines indicate a topological phase transition identified by the minimum of the bandgap. }
    \label{fig:PlPr_compare_nogauss}
\end{figure}

The quantity $\mathcal{P}_l^*\mathcal{P}_r$, indicative of non-local correlations, is presented in Fig.~\ref{Fig_disquasi1d_1} (d). This is zero when $0\leq h/\Delta\lesssim 3$, but acquires significantly large magnitudes \textit{before} the {true} topological phase transition ($3\lesssim h/\Delta\lesssim h_c$). This is due to the presence of partially separated Andreev bound states (or the quasi-Majoranas), demonstrated in  Fig.~\ref{Fig_disquasi1d_1} (e)-ii, and (d)-iii. Here, one Majorana mode is localized at one end of the system, and the other mode is only partially separated from the first one. When the magnetic field is further increased, one transitions to the topological phase with two MBSs at the two ends of the system (Fig.~\ref{Fig_disquasi1d_1} (e)-iv, (e)-v, and (e)-vi). The behavior of the quantity $\mathcal{P}_l^*\mathcal{P}_r$ in the topological region is similar to the case of quasi-Majoranas. In Fig.~\ref{fig:plstarprphase} we plot $\mathcal{P}_l^*\mathcal{P}_r$ in the parameter space of $\mu$ and $h$ for both the clean and the disordered system. The plot reiterates the earlier observations of Fig.~\ref{Fig_disquasi1d_1}, i.e., $\mathcal{P}_l^*\mathcal{P}_r$ showing significant deviation from -1 in the topological region, and also being significantly close to -1 in the non-topological region.
Finally, we examine the quantity $\mathcal{P}_l^*\mathcal{P}_r$ for just the lowest-energy mode without the Dirac-Delta function weighting introduced in Eq.~\ref{Eq:P_region2}. This is plotted in Fig.~\ref{fig:PlPr_compare_nogauss} (b). While this showed no oscillations for the clean system (Fig.~\ref{fig:PlPr_compare_nogauss} (a)), oscillations are still present both in the trivial and topological phase for the disordered case.


\section{Summary and Conclusions}
We presented the behavior of Majorana polarization in two different disordered systems: a chiral-symmetric one-dimensional (1d) semiconducting nanowire and a quasi-one-dimensional system that breaks chiral symmetry, both subject to Rashba spin-orbit coupling, in-plane Zeeman field, and proximity-induced superconductivity. Our analysis reveals that Majorana polarization \textit{alone} cannot always reliably distinguish a true topological Majorana bound state from the quasi Majorana modes (or psABS) in either case. 
Topological MBS must satisfy the condition $\mathcal{P}^*_l \cdot \mathcal{P}_r \sim -1$,  where $\mathcal{P}_l$ and $\mathcal{P}_r$ are the polarizations of the left and right halves of the system. Two additional conditions must be satisfied as well for the identification of topological MBS and their applicability in topological quantum computation: the presence of a topological bandgap and the localization of wavefunctions at the edges.
We identified scenarios involving partially separated Andreev bound states in the trivial phase, where the condition $\mathcal{P}_r\cdot\mathcal{P}^*_l \sim -1$ is satisfied, yet the criteria of a finite topological bulk gap and edge-localized wavefunctions are not met. Furthermore, we identified scenarios where $\mathcal{P}_r\cdot\mathcal{P}^*_l \sim 0$  even in the topological phase. Our conclusions remain robust irrespective of the chosen definition of Majorana polarization, whether based on the chiral or the particle-hole framework. 
These observations suggest that Majorana polarization as a measure of topologically protected Majorana modes useful for TQC, is not always a definitive indicator. The robust identification of topological Majorana bound states (MBS) for applications in topological quantum computation (TQC) necessitates a combination of complementary diagnostics, such as polarization, energy spectra, and wavefunction localization.
Our work will find relevance in current and upcoming studies on  Majorana fermions in solid-state heterostructures.

\textit{Acknowledgment:} G.S. was funded by ANRF-SERB Core Research Grant CRG/2023/005628. S.K. was funded by IIT Mandi HTRA fellowship. S.T. acknowledges ARO Grant W911NF2210247. We thank Nick Sedlmayr for useful discussions.\\



\bibliography{biblio.bib}
\end{document}